Energetics and electronic structure of native point defects in $\alpha$-Ga$_2$O$_3$

Takuma Kobayashi[1], Tomoya Gake[1], Yu Kumagai[1], Fumiyasu Oba[1], and Yu-ichiro Matsushita[1]

[1]Laboratory for Materials and Structures, Institute of Innovative Research, Tokyo Institute of Technology, Yokohama 226-8503, Japan

We report first-principles calculations that clarify the formation energies and charge transition levels of native point defects (Ga and O vacancies, interstitials, and a Ga vacancy-O vacancy pair) in corundum structured $\alpha$-Ga$_2$O$_3$. Either under a Ga- or O-rich growth condition, the negatively-charged Ga vacancy and the positively-charged Ga interstitial on a site surrounded by six O atoms are dominant when the Fermi level approaches the conduction and valence band edges, respectively. These defects would compensate carrier electrons and holes, respectively. Ga-rich conditions relatively suppress the formation of the Ga vacancy and, therefore, are suited for extrinsic $n$-type doping of $\alpha$-Ga$_2$O$_3$.

Gallium oxide ($Ga_2O_3$) has emerged as an attracting material for a wide range of applications[1-3] such as power electronics[4,5] and deep-ultraviolet photodetectors[6,7]. Among the five common polymorphs in $Ga_2O_3$ (i.e., $\alpha$, $\beta$, $\gamma$, $\delta$, and $\varepsilon$),[8] the monoclinic structured $\beta$-$Ga_2O_3$ has been intensively studied and a number of devices were demonstrated in the past decades. For instance, Schottky barrier diodes (SBD),[9] metal-semiconductor field effect transistors (MESFETs)[10] and depletion-mode metal-oxide-semiconductor field effect transistors[11] were reported. Many of theoretical studies have also been focusing on the fundamental and defect properties of $\beta$-$Ga_2O_3$ so far.[12-19] For instance, formation energies and charge transition levels were reported for its native point defects including vacancies of Ga and O ($V_{Ga}$ and $V_O$, respectively) and interstitials ($Ga_i$ and $O_i$),[15-17] and donor-[18] and acceptor-type impurities[19]. It was shown that $V_O$ behaves as a deep donor,[15-17] while impurities such as Sn, Si, Ge, F, and Cl act as shallow donors.[18]

In recent years, corundum structured $\alpha$-$Ga_2O_3$ has also been attractive as well. Starting with the heteroepitaxy of $\alpha$-$Ga_2O_3$ thin films on $\alpha$-$Al_2O_3$ substrates by ultrasonic mist chemical vapor deposition,[20] fabrication and operation of devices, such as SBDs[21] and MESFETs[22], were reported. Its band structure,[23,24] interfacial band alignment,[25] thermodynamic properties,[12] and hole polarons[24] have also been investigated using first-principles calculations. However, only few theoretical reports on point defect properties are available for $\alpha$-$Ga_2O_3$; although there is a study concerning impurity properties (Si, Sn, and Mg),[26] systematic investigation into the properties of native defects such as vacancies and interstitials is still lacking and thus are highly demanded.

In this study, we investigated defect properties of $\alpha$-$Ga_2O_3$ by first-principles calculations using a hybrid functional. We calculated the formation energies and charge transition levels under Ga- or O-rich crystal growth conditions, focusing on the native point defects.

All calculations in this study were performed based on projector augmented wave (PAW) method[27] as implemented in the Vienna Ab-initio Simulation Package (VASP) code.[28,29] We applied the Heyd-Scuseria-Ernzerhof (HSE) hybrid functional[30-32] with Fock-exchange mixing and screening parameters of 0.35 and 0.208 Å$^{-1}$, respectively. The parameters were determined so as to reproduce the experimental bandgap of $\beta$-Ga$_2$O$_3$[13,18] as described in detail later. The cutoff energies in the plane wave basis set were set to 520 and 400 eV for the calculations of perfect $\alpha$-Ga$_2$O$_3$ and defect systems, respectively. First, the lattice parameters were determined by relaxing the volume of the 10-atom primitive cell. 4×4×4 $k$ points were sampled during this calculation, and the optimization was performed until the residual forces acting on the atoms were less than 10 meV/Å. Then, the optimization of atomic positions was performed in the defect system to obtain stable defect configurations. A 120-atom supercell and 2×2×2 $k$ points were used in investigating the defects. The geometry optimization was performed until the residual forces were less than 50 meV/Å. Spin polarization of the defects was considered in these defect calculations.

The formation energy $E_F$ of defect $D$ in charge state $q$ ($D^q$) is given as[33-35]

$$E_F[D^q] = (E[D^q] + E_C[D^q]) - E_P - \sum_i n_i \mu_i + q(\varepsilon_{VBM} + \Delta\varepsilon_F). \qquad (1)$$

Here, $E[D^q]$, $E_P$, $\varepsilon_{VBM}$, and $\Delta\varepsilon_F$ are the total energy of the supercell with $D^q$, that of the perfect crystal supercell, the energy level of the valence band maximum (VBM), and the Fermi level with respect to the VBM, respectively. $n_i$ and $\mu_i$ are the number of added ($n_i > 0$) or removed ($n_i < 0$) $i$-type atom and its chemical potential, respectively. $E_C[D^q]$ is a correction term for removing the spurious long-range Coulomb interactions between $D^q$, its periodic images, and the background charge under three-dimensional periodic boundary conditions. We applied the extended Freysoldt-Neugebauer-Van de Walle (FNV) scheme[33,36] for the correction, which can correct energies of charged defects accurately in various systems.[33,36-38] We considered two extreme crystal growth conditions, where the chemical potentials of the atomic species are given as

$$\mu_{Ga} = E(Ga), \ \mu_O = \frac{1}{3}E(Ga_2O_3) - \frac{2}{3}E(Ga) \quad \text{(Ga-rich)}, \tag{2}$$

$$\mu_{Ga} = \frac{1}{2}E(Ga_2O_3) - \frac{3}{4}E(O_2), \ \mu_O = \frac{1}{2}E(O_2) \quad \text{(O-rich)}, \tag{3}$$

where $E(Ga)$, $E(Ga_2O_3)$, and $E(O_2)$ are the total energies of $\alpha$-Ga, $\alpha$-Ga$_2$O$_3$, and an O$_2$ molecule per formula unit, respectively. Since the defect formation energies depend linearly on atomic chemical potentials as given in Eq. (1), the results for growth conditions between these limits can be readily estimated by interpolation.

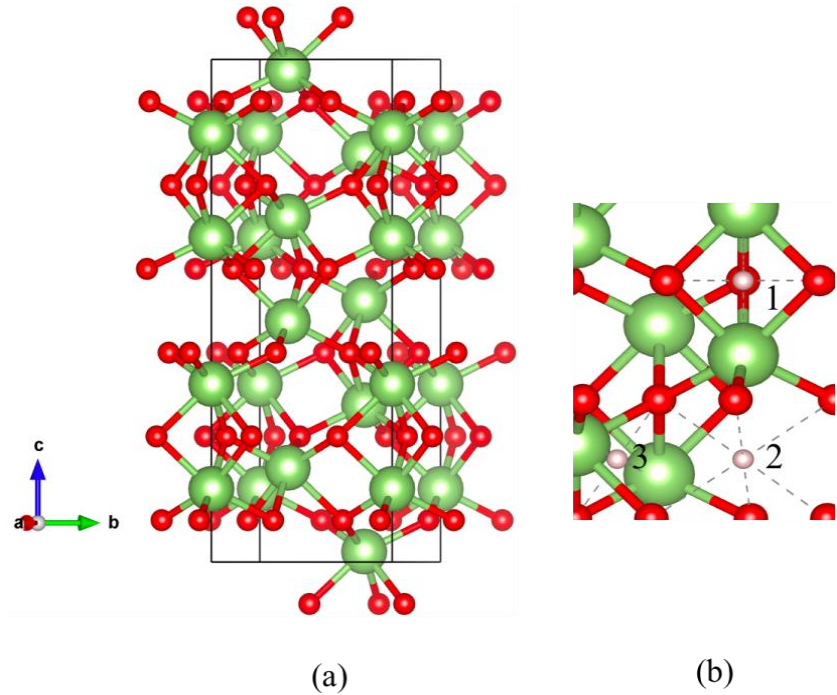

(a)  (b)

Fig. 1: Crystal structure of $\alpha$-Ga$_2$O$_3$: (a) An optimized 30-atom conventional cell and (b) an enlarged view describing the three inequivalent interstitial sites, labeled 1, 2, and 3. The green, red, and small white balls describe the Ga and O atoms, and the interstitial sites, respectively.

Figure 1 (a) shows a 30-atom conventional cell of $\alpha$-Ga$_2$O$_3$ lattice. The optimized lattice constants are 4.964 and 13.392 Å for the *a*- and *c*-axes, respectively, which agree well with the experimental values of 4.983 and 13.433 Å[39] at room temperature. All the Ga and O atoms in $\alpha$-Ga$_2$O$_3$ are symmetrically equivalent, respectively, so we investigated the vacancies ($V_{Ga}$ and $V_O$) on these sites. For a Ga vacancy-O vacancy pair ($V_{Ga}$-$V_O$), we only considered the nearest neighbor case. For the interstitials (Ga$_i$ and O$_i$), we investigated three inequivalent high-symmetry interstitial sites as depicted in Fig. 1 (b): A site surrounded by two Ga and three O atoms (site 1), six O atoms (site 2), and two Ga and two O atoms (site 3).

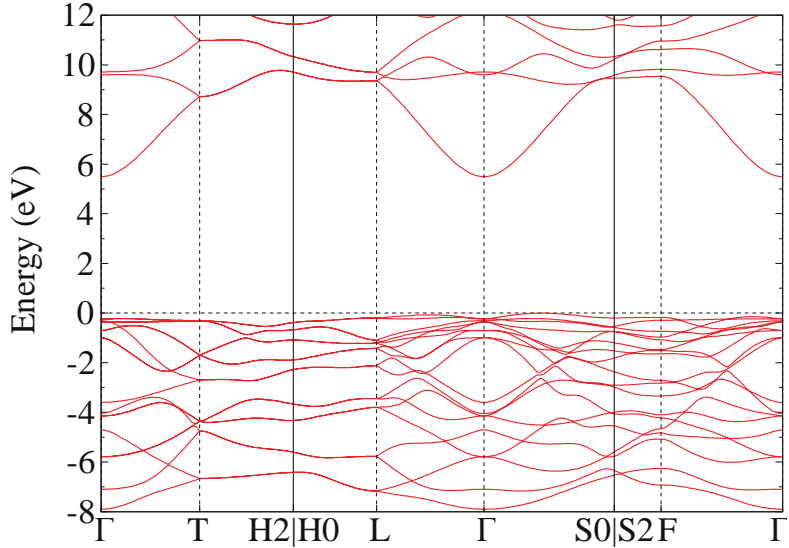

Fig. 2: Calculated band diagram of $\alpha$-Ga$_2$O$_3$. The conduction band and valence band edges are located at the $\Gamma$ point and a point between the $\Gamma$ and S0 points, respectively. The zero of the energy is set at the VBM. The band paths conform to Ref. 40.

The calculated band structure of $\alpha$-Ga$_2$O$_3$ is depicted in Fig. 2. Since the experimental bandgap of $\alpha$-Ga$_2$O$_3$ varies in the literature (4.9 - 5.6 eV[20,41-44]), we determined the Fock-exchange mixing parameter in the HSE functional to be 0.35 with the screening parameter kept fixed at a standard value of 0.208 Å$^{-1}$ [45]) so that the calculated bandgap of $\beta$-Ga$_2$O$_3$ (4.89 eV) agrees well with the experimental one (4.9 eV[46])). This approach has been shown to reproduce the experimental ionization potential well and nearly satisfies generalized Koopmans' theorem for a self-trapped hole in $\beta$-Ga$_2$O$_3$.[24]) The bandgap of $\alpha$-Ga$_2$O$_3$ obtained using this functional is 5.49 eV in the indirect-type band structure. This value is close to a previously reported indirect gap of 5.39 eV from a G$_0$W$_0$@HSE03 calculation.[23])

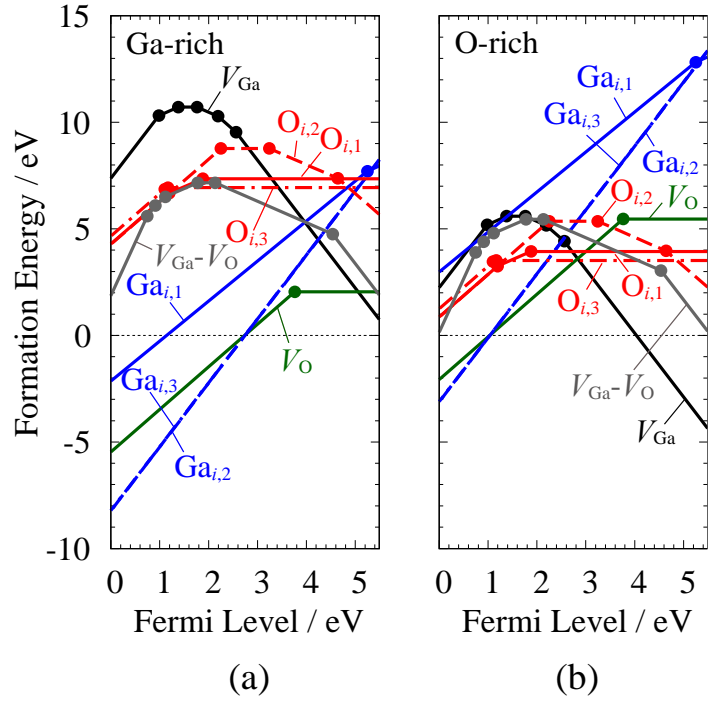

Fig. 3: Formation energies of Ga and O vacancies ($V_{Ga}$ and $V_O$), interstitials (Ga$_i$ and O$_i$), and a Ga vacancy-O vacancy pair ($V_{Ga}$-$V_O$) in $\alpha$-Ga$_2$O$_3$ either at (a) the Ga-rich and (b) the O-rich limit. The zero of the Fermi level is set at the VBM. For the interstitials, results for the defects at interstitial sites 1, 2, and 3 (see Fig. 1 (b)) are separately shown.

Figure 3 shows the formation energies of the investigated defects under either Ga- or O-rich condition. Either under Ga- or O-rich condition, negatively-charged $V_{Ga}$ and positively-charged $Ga_i$ are energetically favorable when the Fermi level approaches the CBM and VBM, respectively. The formation energy of $Ga_i$ takes a negative value under either Ga- or O-rich conditions when the Fermi level is low. This implies that the realization of a *p*-type material is difficult because of the carrier compensation by $Ga_i$, as well as the small hole polaron formation and related deep, polaronic nature of dopant-induced acceptor states.[24] At the Ga-rich limit, neutral and positively-charged $V_O$ are stable in a wide range of the Fermi level position from intrinsic to *n*-type conditions. However, the O vacancy is a deep donor with a transition level far below the CBM, and unlikely to be a source of native *n*-type conductivity. At the O-rich limit, positively-charged $V_O$, as well as $Ga_i$, is rather stable for lower Fermi level values. We also see that the formation energy of $V_{Ga}$ takes a negative value when the Fermi level is near CBM at the O-rich limit, indicating that even *n*-type doping would be difficult for the extreme O-rich condition. Thus, crystal growth under a condition close to the Ga-rich limit is preferred in realizing *n*-type material.

We found that $V_{Ga}$ takes −3 - +3 charge states depending on the position of the Fermi level. In α-$Ga_2O_3$, $V_{Ga}$ is surrounded by six O atoms. In the neutral charge state, three holes localize onto 2p orbitals of three different O atoms around $V_{Ga}$, exhibiting polaronic charge localization, and the remaining three O atoms can capture three more holes. Thus, $V_{Ga}$ acts as a triple donor as well as a triple acceptor depending on the Fermi level. Similar hole localization is reported in other oxide semiconductors such as ZnO[47]. We found that $V_O$ acts as a deep donor showing small negative-*U* behavior; $U = E_F[V_O^0] + E_F[V_O^{+2}] - 2E_F[V_O^{+1}] = -0.16$ eV. The energy gain for the formation of a Ga vacancy-O vacancy pair (i.e., $V_{Ga} + V_O \rightarrow V_{Ga}$-$V_O$) is about 0.9 eV at the *n*-type limit, indicating the vacancy pair formation in *n*-type materials, in particular at low temperatures where entropic energy gain is small for isolated defects.

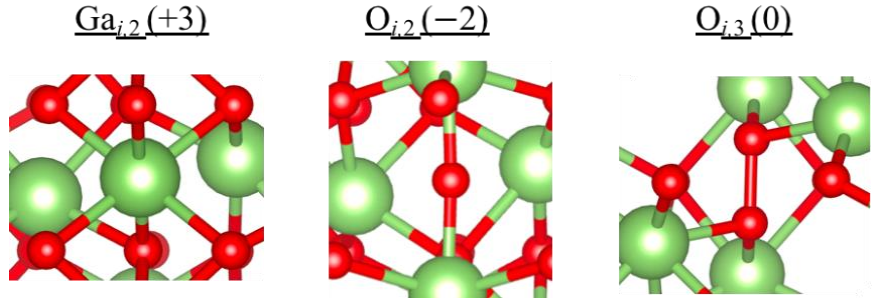

Fig. 4: Optimized structure of the stable interstitials ($Ga_i$ and $O_i$): $Ga_i$ at the interstitial site 2 ($Ga_{i,2}$) in Fig. 1(b) and $O_i$ at the interstitial sites 2 ($O_{i,2}$) and 3 ($O_{i,3}$). $Ga_{i,2}$, $O_{i,2}$, and $O_{i,3}$ are in the +3, −2, and neutral charge states, respectively. The color code is the same as in Fig. 1.

The optimized structures of the stable Ga and O interstitials are shown in Fig. 4. The most stable form of $Ga_i$ is $Ga_{i,2}$, in which the interstitial Ga atom is surrounded by six O atoms. Note that the $Ga_{i,3}$ eventually converged to the same structure as $Ga_{i,2}$, and thus shows similar formation energy as $Ga_{i,2}$ (Fig. 3). In Fig. 3, we see that $Ga_{i,2}$ is always in +3 charge state regardless of the position of the Fermi level, indicating that the defect supplies electrons to the conduction band if they are not compensated by acceptor-type defects and thus operates as a shallow donor.

Among the oxygen interstitials, we found that the split interstitials, $O_{i,1}$ and $O_{i,3}$, are stable when the Fermi level is located at the lower half of the bandgap, whereas the negatively-charged interstitial without splitting, $O_{i,2}$, becomes stable under *n*-type conditions. The reason why the split interstitials ($O_{i,1}$ and $O_{i,3}$) are not stabilized in negatively-charged states can be understood by considering the molecular orbital of an $O_2$ molecule; In the ground state of an isolated, neutral $O_2$ molecule, two up-spin electrons are in the antibonding π* states, realizing a triplet state. In $Ga_2O_3$, each O atom formally receives two electrons from Ga atoms because of the difference of electronegativity of Ga and O atoms, and thus the split interstitials ($O_{i,1}$ and $O_{i,3}$) together with a host O atom accommodates two more electrons than a neutral $O_2$ molecule. Then, the π* states are fully occupied with 4 electrons in the neutral condition. To add an additional electron into this system, the electron should go into σ* states with high energy loss. Thus, the split interstitials do not take negatively-charged states. Indeed, we confirmed that 2 up- and 2 down-spin electrons fully occupy the states with similar energy levels ($E_V$ + 0.55 eV and $E_V$ + 0.58 eV) in the neutral charge state of $O_{i,3}$. The fact that the bond-length of two oxygen atoms (1.372 Å) in the neutral split-interstitial is longer than that of the calculated value for an $O_2$ molecule (1.207 Å) also suggests that electrons get into the antibonding π* states in the case of the split-interstitial in the neutral charge state.

In summary, we report the formation energies and charge transition levels of the native point defects in corundum-structured $α$-$Ga_2O_3$. We found that negatively-charged $V_{Ga}$ and positively-charged $Ga_i$ are stable when the Fermi level is near the CBM and VBM, respectively, regardless of crystal growth conditions (Ga-rich or O-rich). Either under Ga- or O-rich condition, the formation energy of $Ga_i$ takes a negative value when the Fermi level approaches the VBM. This implies that the realization of a *p*-type material is difficult because of a strong carrier compensation. At the Ga-rich limit, neutral and positively-charged $V_O$ are stable within a wide range of the Fermi level from an intrinsic to n-type condition. However, the O vacancy is a deep donor and unlikely to be a source of native *n*-type conductivity. At the O-rich limit, positively-charged $V_O$ becomes comparable in energy to $Ga_i$ at lower Fermi levels. We also found that the formation energy of $V_{Ga}$ takes a negative value when the Fermi level approaches the CBM at the O-rich limit, indicating that even extrinsic *n*-type doping is difficult in thermal equilibrium state for such O-rich conditions. Thus, crystal growth under a Ga-rich condition is preferred in *n*-type doping.

Computations were performed mainly at the Center for Computational Science, University of Tsukuba, and the Supercomputer Center at the Institute for Solid State Physics, The University of Tokyo. The authors acknowledge the support from JSPS Grant-in-Aid for Scientific Research (A) (Grant Number: 18H03770 and 18H03873).